# Observations on the motion of a Tachyon

**Chandru Iyer**[1]

[1]Techink Industries, C-42 Phase II, Noida, India 201 305
 Email : chandru.iyer@luxoroffice.com

*Abstract: Some aspects of the motion of a tachyon is discussed. It is shown that the inertial frame Σ ' around which the tachyon switches the direction of its motion, does not observe any movement of the tachyon at all. Inertial frames on either side of Σ ' observe the tachyon to be moving at very large speeds but in opposite direction. Σ ' itself observes only a sudden appearance and immediate disappearance of a long rod like object. Thus unbounded speeds in either direction give the same result in the limit. This suggests that negative numbers as a physical quantity are not meaningful. Subtraction can be used integral to a formula but the final result has to be interpreted with a positive answer. This means the abstract quantity  –∞ indicating an unbounded negative number is not meaningful. The situation is also compared with Tan (π/2)$^+$ and Tan(π/2)$^-$. The conclusion is that in the limit, travel at unbounded speed is direction independent and gives the connotation of many identities to the same particle.*

Tachyons (particles which travel faster than light) arise naturally in superstring theory; a detailed analysis from algebraic as well as space-time diagram viewpoints is given in [1]. For objects which travel below the speed of light as observed by an inertial frame, the time order of two points (events) on the world line of the object remains unaltered as observed by any other inertial frame. For tachyons, this is not the case.  If a tachyon is traveling at u (>c) with respect to an inertial frame Σ, then for an inertial frame Σ ' which is traveling with respect to Σ at a speed greater than $c^2/u$, the time order of two points on the tachyon's trajectory reverses and thus the tachyon is observed to travel in the opposite direction.

Let us consider two points (events) in the space-time continuum with coordinates $(x_1, t_1)$ and $(x_2, t_2)$ as observed by an inertial frame Σ and let $\Delta x = x_2 – x_1$, and $\Delta t = t_2 – t_1$. As long as $|\Delta x/ \Delta t | \leq c$, the time order of these two events do not change with respect to any inertial frame Σ ' moving at V with respect to Σ  with $|V| < c$. When $|\Delta x/ \Delta t | = c$ , $|\Delta x'/ \Delta t'|$ remains c as observed by Σ ' (second postulate of 'special relativity') and only in this case the both the sign and the magnitude of $\Delta x'/ \Delta t'$ remains unaffected as observed by any arbitrary inertial frame. .

Two points on the world line of the tachyon, represent the third case when $|\Delta x/ \Delta t | >$ c. For this case, $|\Delta x'/ \Delta t'|$ continues to remain >c as observed by Σ '. But when V= $c^2 \Delta t / \Delta x$ (which is less than c in absolute value for this case), we have a situation that the velocity of the tachyon is ± ∝.

The point in question is can the tachyon travel at infinite velocity with respect to the inertial frame Σ ', which is a real inertial frame under SR. Even if it does, which direction the tachyon is traveling is also not clear. We can resolve this by considering the fact that under SR each event is an entity by itself and there is no connection or continuity between them. When a particle travels at a speed of $+\infty$, what $\Sigma'$ will observe is the instantaneous appearance and immediate disappearance of a very long rod like linear object. This is because the particle appears simultaneously at all points along the spatial dimension. The curious thing is the same observation will take place even if the particle is traveling at $-\infty$. In that sense a particle traveling at $+\infty$ or $-\infty$ is indistinguishable. On either side of $V = c^2 \Delta t / \Delta x$, the tachyon's velocity is positive and negative but greater than c in absolute value.

The space-time imprint of the tachyon can be compared with the propagation of the collision point as in the collision of two parallel rods in relative motion with the line of motion at an inclination with the axis of the rods[2]

The observation of the appearance and immediate disappearance of a long rod like object by $\Sigma'$ will be interpreted by $\Sigma$ as follows : "observers in $\Sigma'$ had asynchronous clocks and due to this they observed a moving tachyon's positions at various times as simultaneous". However, under the first postulate of SR, the observations of $\Sigma'$ are as valid as those of $\Sigma$ in their respective frames.

**Discussion:** Large negative numbers have to be interpreted as positive numbers with a change of direction. In an isotropic environment this will give the same result as a large positive number in the original direction. Thus the observation by $\Sigma'$ is of neither a particle moving at $+\infty$ or $-\infty$ but a sudden appearance and immediate disappearance of a rod like object.

This situation may be compared with $\tan(\pi/2)$. The left limit tends to $-\infty$ and the right limit tends to $+\infty$. However, the line that subtends $\pi/2$ with the x axis is well defined and it is the y axis. If we consider that physically $-\infty$ and $+\infty$ will lead to the same situation, the result in both the cases (the tachyon with respect to $\Sigma'$ and $\tan(\pi/2)$) is not surprising at all and it arises out of isotropy of space. Especially in the case of the tachyon, while all other inertial frames observed a moving tachyon, $\Sigma'$ observed a rod that appears momentarily and this is very interesting.

**Conclusion:** Travel, by definition, means a finite distance covered in a finite non-zero time interval. In the limit travel at infinite speed is direction independent. Travel at infinite speed gives the connotation of many identities to the same particle. Therefore the observation by a particular inertial frame that the tachyon traveled at infinite speed is in variance with the observation of a finite speed (of the tachyon) by all other inertial frames.